\documentclass[
    ,final            
  ]
  {aipproc}

\layoutstyle{8x11single}

\begin{document}

\title{Charmed mesons in nuclear matter}


\author{L. Tolos}{
  address={Theory Group. KVI. University of Groningen,
Zernikelaan 25, 9747 AA Groningen, The Netherlands}
}

\author{D. Gamermann}{
  address={Instituto de F{\'\i}sica Corpuscular (centro mixto CSIC-UV)
Institutos de Investigaci\'on de Paterna, Aptdo. 22085, 46071, Valencia, Spain}
}

\author{C. Garcia-Recio}{
  address={Departamento de F{\'\i}sica At\'omica, Molecular y Nuclear, 
Universidad de Granada, E-18071 Granada, Spain}
}

\author{E. Oset}{
  address={Instituto de F{\'\i}sica Corpuscular (centro mixto CSIC-UV)
Institutos de Investigaci\'on de Paterna, Aptdo. 22085, 46071, Valencia, Spain}
}

\author{R. Molina}{
  address={Instituto de F{\'\i}sica Corpuscular (centro mixto CSIC-UV)
Institutos de Investigaci\'on de Paterna, Aptdo. 22085, 46071, Valencia, Spain}
}

\author{J. Nieves}{
  address={Instituto de F{\'\i}sica Corpuscular (centro mixto CSIC-UV)
Institutos de Investigaci\'on de Paterna, Aptdo. 22085, 46071, Valencia, Spain}
}

\author{A. Ramos}{
address={Departament d'Estructura i Constituents de la Mat\`eria,
Universitat de Barcelona,
Diagonal 647, 08028 Barcelona, Spain}
}

\begin{abstract}
We obtain the properties of charmed mesons in dense matter using a coupled-channel approach which accounts for Pauli blocking effects and meson self-energies in a self-consistent manner. We study the behaviour of dynamically-generated baryonic resonances together with the open-charm meson spectral functions in this dense nuclear environment. We discuss the implications of the in-medium properties of open-charm mesons on the  $D_{s0}(2317)$ and the predicted $X(3700)$ scalar resonances, and on the formation of $D$-mesic nuclei.
\end{abstract}

\maketitle


\section{Introduction}
The interest on the properties of open and hidden charm mesons was initiated in the context of heavy-ion collisions in connection to the charmonium suppression
\cite{matsui86} as a probe for the formation of Quark-Gluon Plasma. Recently charmed baryonic
resonances have received  a lot of attention motivated by the
discovery of quite a few new states by the CLEO, Belle and BABAR
collaborations \cite{facility00}. Moreover, the future FAIR facility
at GSI \cite{gsi00} will move from the light quark sector to the heavy
one and will face new challenges where charm plays a dominant role.
The CBM (Compressed Baryonic Matter) experiment at FAIR/GSI 
will extend the GSI programme for in-medium modification of hadrons in the
light quark sector, and provide first insight into the charm-nucleus
interaction. Therefore, the modifications of the properties of open and hidden charm mesons in a hot and dense environment are being the focuss of recent studies.

The in-medium modification of the properties of open-charm mesons ($D$ and $\bar D)$ may help to explain the $J/\Psi$ suppression in a hadronic environment as well as the possible formation of $D$-mesic nuclei. Moreover, changes in the properties of open-charm mesons will affect the renormalization of charm and hidden charm scalar meson resonances in nuclear matter, providing information about their nature.

In the present paper we obtain the properties of open-charm mesons in dense matter within a self-consistent approach in coupled channels. We study the behaviour of dynamically generated charmed baryonic resonances as well as the open-charm meson spectral functions in this dense medium. We then analyze the effect of the self-energy of $D$ mesons on the properties of dynamically-generated charm and hidden charm scalar resonances, such as the $D_{s0}(2317)$ and the predicted $X(3700)$ resonances. We finally provide some recent results on $D$-nucleus bound states.

\section{Open-charm mesons in dense nuclear matter}
The self-energy and, hence, the spectral function for open-charm ($D$ and $\bar D$) mesons is obtained following a self-consistent coupled-channel procedure. The transition potential of the Bethe-Salpeter equation or $T$-matrix ($T$) is derived from effective lagrangians, which will be discussed in the following. The self-energy is then obtained summing the transition amplitude $T$ for the different isospins over the nucleon Fermi distribution at a given temperature, $n(\vec{p},T)$: 
\begin{eqnarray}
\Pi(q_0,{\vec q},T)= \int \frac{d^3p}{(2\pi)^3}\, n(\vec{p},T)  \, [\, {T}^{(I=0)} (P_0,\vec{P},T) +
3 \, {T}^{(I=1)} (P_0,\vec{P},T)\, ]\ , \label{eq:selfd}
\end{eqnarray}
\noindent
where $P_0=q_0+E_N(\vec{p},T)$ and $\vec{P}=\vec{q}+\vec{p}$ are
the total energy and momentum of the meson-nucleon pair in the nuclear
matter rest frame, and ($q_0$,$\vec{q}\,$) and ($E_N$,$\vec{p}$\,) stand  for
the energy and momentum of the meson and nucleon, respectively, also in this
frame. The self-energy must be determined self-consistently since it is obtained from the
in-medium amplitude $T$ which contains the meson-baryon loop function, and this last quantity itself
is a function of the self-energy. Then, the meson spectral function  reads
\begin{eqnarray}
S(q_0,{\vec q}, T)= -\frac{1}{\pi}\frac{{\rm Im}\, \Pi(q_0,\vec{q},T)}{\mid
q_0^2-\vec{q}\,^2-m^2- \Pi(q_0,\vec{q},T) \mid^2 } \ .
\label{eq:spec}
\end{eqnarray}

\begin{figure}
\includegraphics[width=0.45\textwidth, height=7cm]{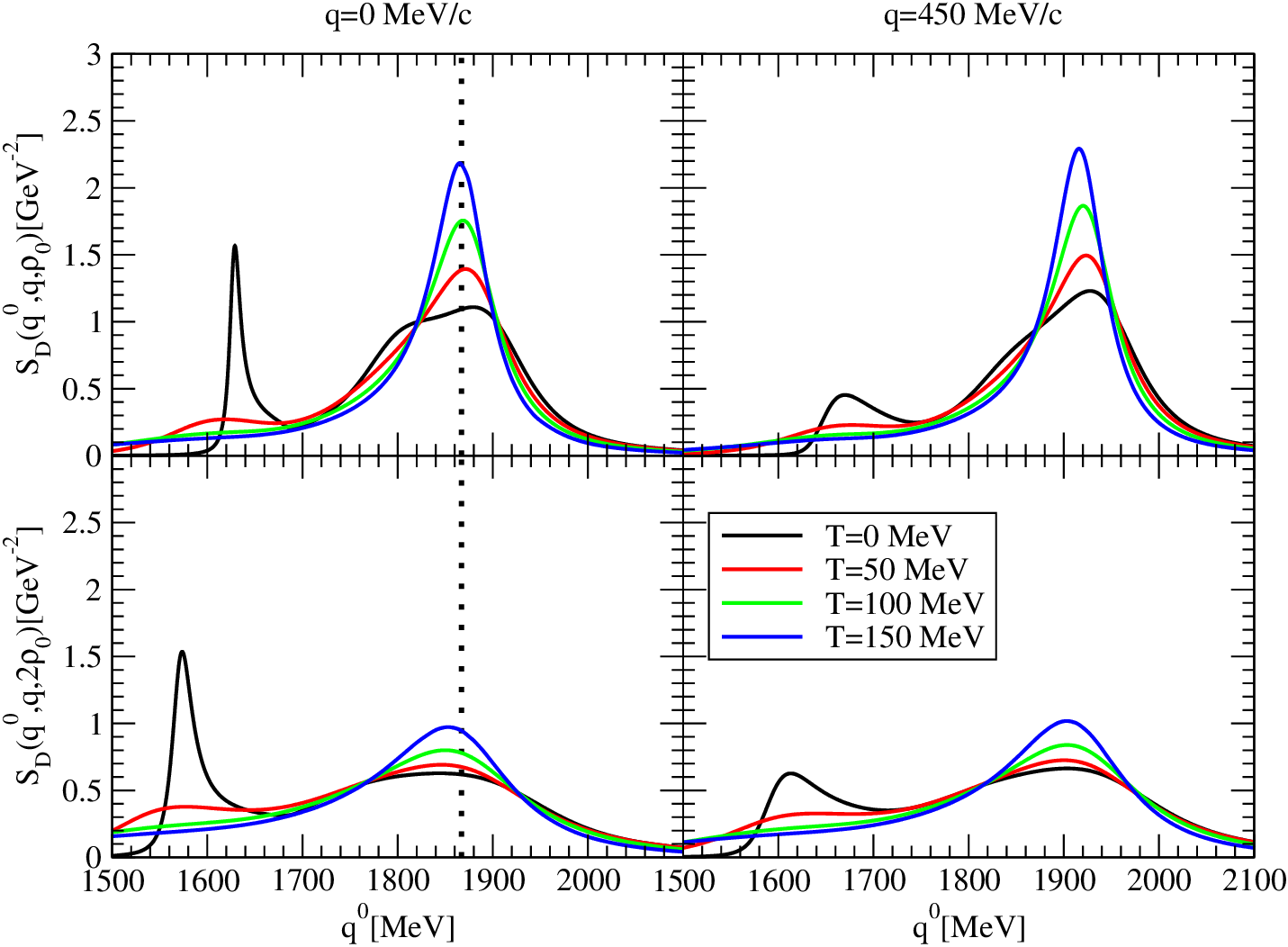}
\caption{The $D$ meson spectral function for different momenta, temperatures and densities for the SU(4) model. We show the $D$ meson free mass for reference (dotted lines). \label{fig1}}
\end{figure}

\subsection{SU(4) $t$-vector meson exchange models}
\label{su4}

The open-charm meson spectral functions are obtained from the Bethe-Salpeter equation in coupled-channels taking, as bare interaction, a type of broken $SU(4)$ $s$-wave Weinberg-Tomozawa (WT) interaction supplemented by an attractive isoscalar-scalar term and using a cutoff regularization scheme. This cutoff is fixed by generating dynamically the $I=0$ $\Lambda_c(2593)$ resonance. A new resonance in $I=1$ channel, $\Sigma_c(2880)$, is generated \cite{LUT06,mizutani06}. The in-medium solution at finite temperature 
incorporates Pauli blocking, baryon mean-field bindings and $\pi$ and $D$ meson self-energies \cite{TOL07}.

In  Fig.~\ref{fig1} we display the $D$ meson spectral function for different momenta, temperatures and densities. At $T=0$ the spectral function shows two peaks. The $\tilde \Lambda_c N^{-1}$ excitation is seen at a lower energy whereas the second one at higher energy corresponds to the quasi(D)-particle  peak  mixed with  the $\tilde \Sigma_c N^{-1}$ state.  Those structures dilute with increasing temperature while the quasiparticle peak gets closer to its free value becoming narrower, as the self-energy
receives contributions from higher momentum $DN$ pairs where the interaction is weaker.
Finite density results in a broadening of the spectral function because of the increased phase space, as previously observed for the $\bar K$ in dense matter \cite{Tolos:2008di}.

\subsection{SU(8) scheme with heavy-quark symmetry}
\label{su8}
Heavy-quark symmetry (HQS) is a QCD spin-flavor symmetry
that appears when the quark masses, such as the charm mass, become
larger than the typical confinement scale. The spin interactions then vanish for infinitely massive
quarks and heavy hadrons come in doublets (if the spin of the light
degrees of freedom is not zero), which are degenerate in the infinite
quark-mass limit. And this is the case for the $D$ meson and its
vector partner, the $D^*$ meson. 

Therefore we calculate the self-energy and, hence, the spectral function of the $D$ and $D^*$ mesons in nuclear matter  from a  simultaneous self-consistent calculation in coupled channels that incorporates HQS. We extend the WT meson-baryon lagrangian to the $SU(8)$ spin-flavor symmetry group as we include pseudoscalars and vector mesons together with $J=1/2^+$ and $J=3/2^+$ baryons \cite{magas09}, following the steps for $SU(6)$ of Ref.~\cite{GarciaRecio:2005hy}. The $SU(8)$ spin-flavor is, however, strongly broken in nature. So that we take into account mass
breaking effects by adopting the physical hadron masses in the tree
level interactions and in the evaluation of the
kinematical thresholds of different channels, as done in the previous $SU(4)$ models. Moreover, we consider the difference between the weak non-charmed and charmed
pseudoscalar and vector meson decay constants. We also improve on the regularization scheme in nuclear matter going beyond the usual cutoff scheme \cite{tolos09}.

\begin{figure}
\includegraphics[width=0.45\textwidth, height=7cm]{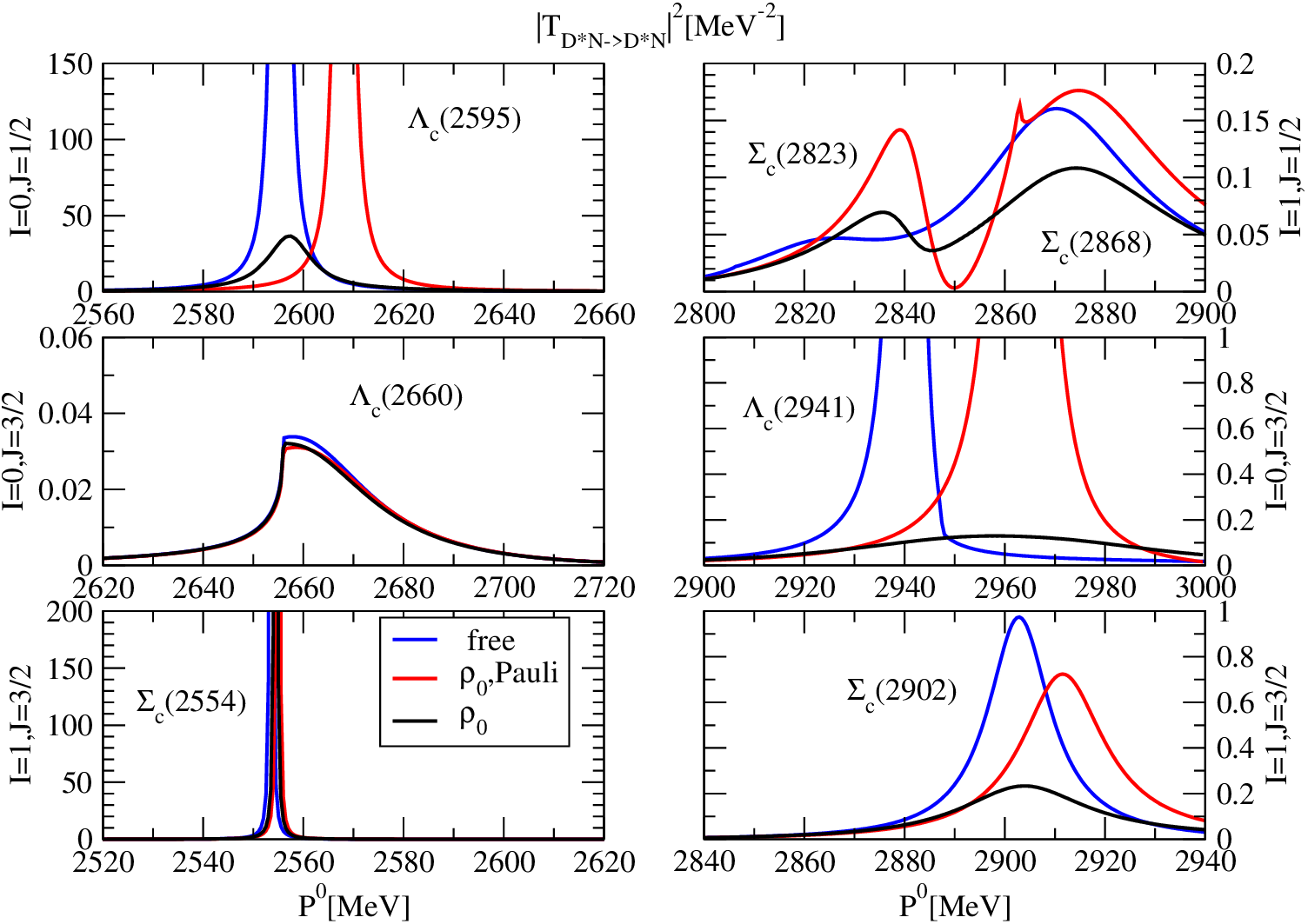}
\hfill
\includegraphics[width=0.45\textwidth, height=5.5cm]{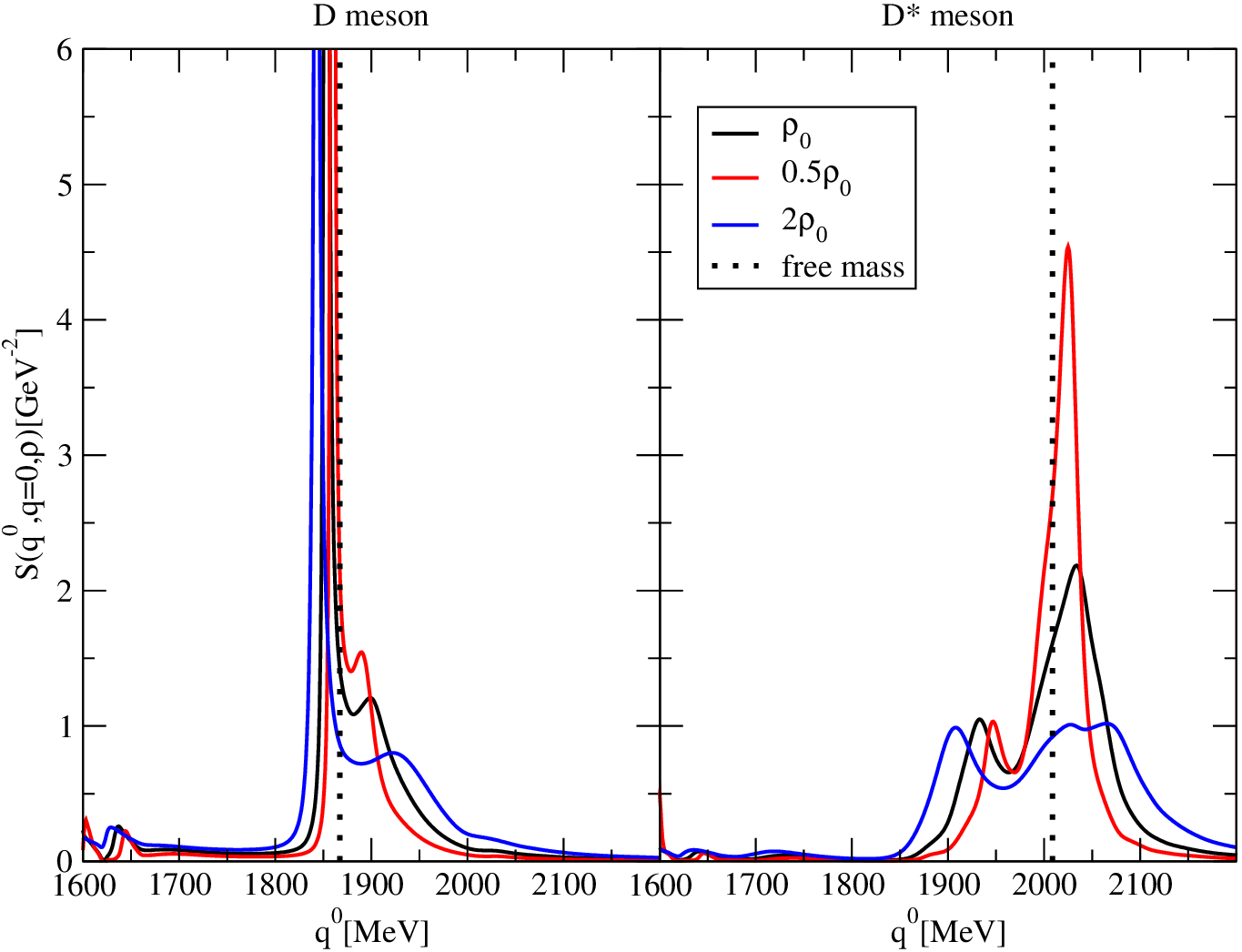}
\caption{Left: Charmed baryonic resonances in nuclear matter in the SU(8) model. Right: $D$ and $D^*$ spectral functions in nuclear matter at $q=0$ MeV/c in the SU(8) scheme. We show the $D$ and $D^*$ meson free masses for reference (dotted lines). \label{fig2}}
\end{figure}


As  seen on the l.h.s. of  Fig.~\ref{fig2}, the $SU(8)$ model generates a wider spectrum of resonances with charm $C=1$
and strangeness $S=0$  compared to the  $SU(4)$ models. While the parameters of both $SU(4)$ and $SU(8)$ models are fixed by the ($I=0$, $J=1/2$) $\Lambda_c(2595)$ resonance, the fact that we incorporate vectors mesons in
the $SU(8)$ scheme generates naturally $J=3/2$ resonances, such as
$\Lambda_c(2660)$, $\Lambda_c(2941)$, $\Sigma_c(2554)$ and
$\Sigma_c(2902)$, some of which might be identified experimentally
\cite{Amsler08}. New resonances are also produced for $J=1/2$, as
$\Sigma_c(2823)$ and $\Sigma_c(2868)$, while others are not obtained in $SU(4)$ models because of the different symmetry breaking pattern used in both models.

The modifications of the mass and width of these resonances in
the nuclear medium are strongly dependent on the coupling to channels
with $D$, $D^*$ and $N$ content, which are modified in the nuclear medium. Moreover, the resonances close to
the $DN$ or $D^*N$ thresholds change their properties more evidently
as compared to those far offshell. The improvement in the
regularization/renormalization procedure of the intermediate
propagators in the nuclear medium beyond the usual cutoff method has
also an important effect on the in-medium changes of the
dynamically-generated resonances, in particular, for those lying far
offshell from their dominant channel, as the case of the
$\Lambda_c(2595)$.

On the r.h.s of  Fig.~\ref{fig2} we display the $D$ and $D^*$ spectral functions, which show then a rich spectrum of resonance-hole states. The $D$ meson quasiparticle peak mixes strongly with $\Sigma_c(2823)N^{-1}$
and $\Sigma_c(2868)N^{-1}$ states while the $\Lambda_c(2595)N^{-1}$ is
clearly visible in the low-energy tail. The $D^*$ spectral function
incorporates the $J=3/2$ resonances, and the quasiparticle peak fully mixes with $\Sigma_c(2902)N^{-1}$ and $\Lambda_c(2941)N^{-1}$. 
As density increases, these $Y_cN^{-1}$ modes tend to smear out and the
spectral functions broaden with increasing phase space.

\section{Scalar resonances in nuclear matter}

The analysis of the properties of scalar resonances in nuclear matter is a valuable tool in order to understand the nature of those states, whether they are $q \bar q$, molecules, mixtures of $q \bar q$ with meson-meson components, or dynamically generated resonances resulting from the interaction of two pseudoscalars. 

We study the charmed resonance $D_{s0}(2317)$ \cite{Kolomeitsev:2003ac,guo06,Gamermann:2006nm} together with a hidden charm scalar meson, $X(3700)$, predicted in Ref.~\cite{Gamermann:2006nm}, which might have been observed by the Belle collaboration \cite{Abe:2007sy} via the reanalysis of Ref.~\cite{Gamermann:2007mu}. Those resonances are generated dynamically solving the coupled-channel Bethe-Salpeter equation for two pseudoscalars \cite{Molina:2008nh}. The kernel is derived from a $SU(4)$ extension of the $SU(3)$ chiral Lagrangian used to generate scalar resonances in the light sector. The $SU(4)$ symmetry is, however, strongly 
 broken, mostly due to the explicit consideration of the masses of the vector 
 mesons exchanged between pseudoscalars \cite{Gamermann:2006nm}. 

The transition amplitude around each resonance for the different coupled channels gives us information about the coupling of this state to a particular channel. The $D_{s0}(2317)$ mainly couples to the $DK$ system, while the hidden charm state $X(3700)$ couples most strongly to $D\bar{D}$. Then, any change in the $D$ meson properties in nuclear matter will have an important effect on these  resonances. Those modifications are given by the $D$ meson self-energy in the $SU(4)$ model without the phenomenological isoscalar-scalar term, but supplemented by the $p$-wave self-energy through the corresponding $Y_cN^{-1}$ excitations \cite{Molina:2008nh}.

\begin{figure}
\includegraphics[width=0.4\textwidth,height=5.5cm]{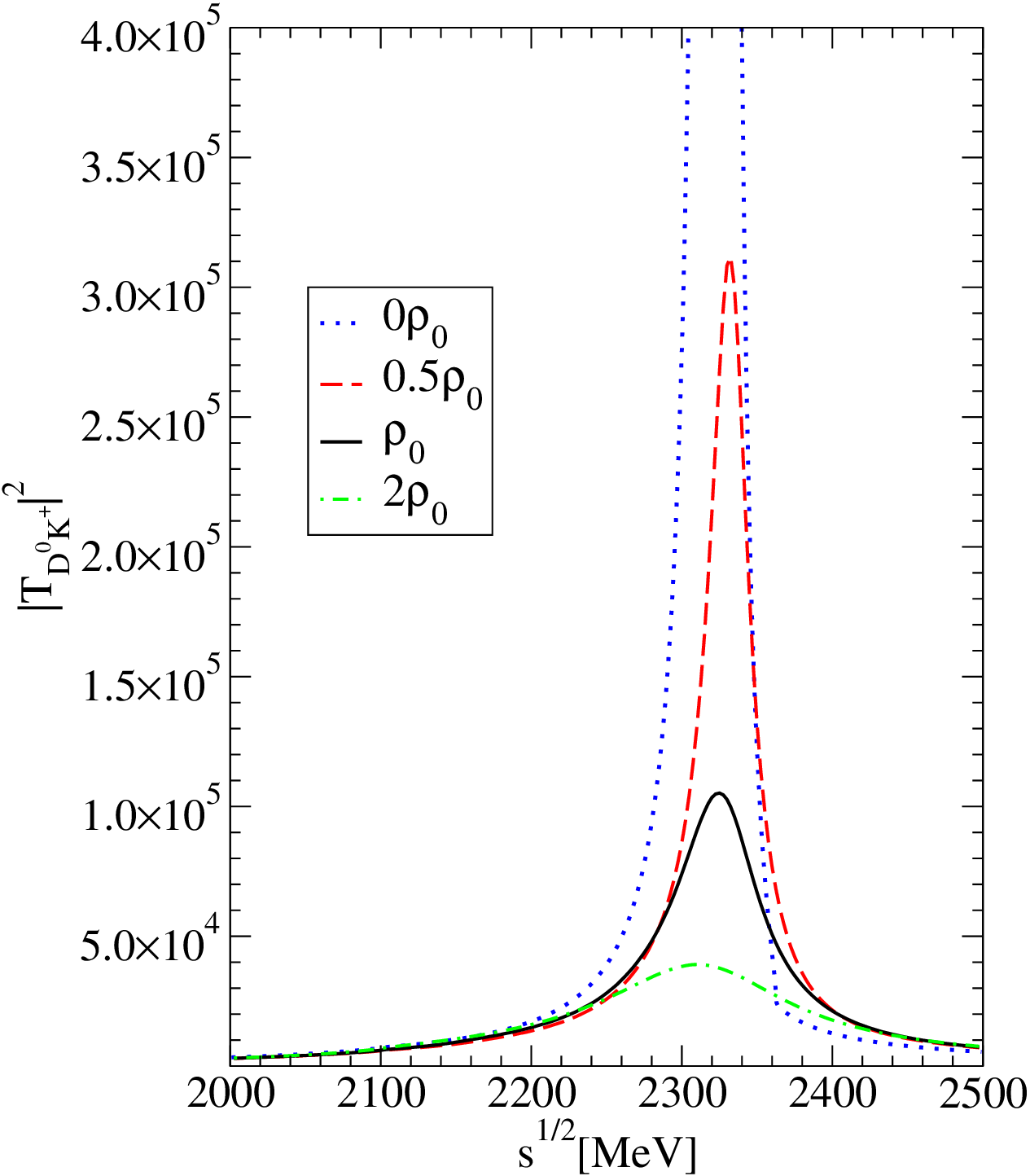}
\hfill
\includegraphics[width=0.4\textwidth,height=5.5cm]{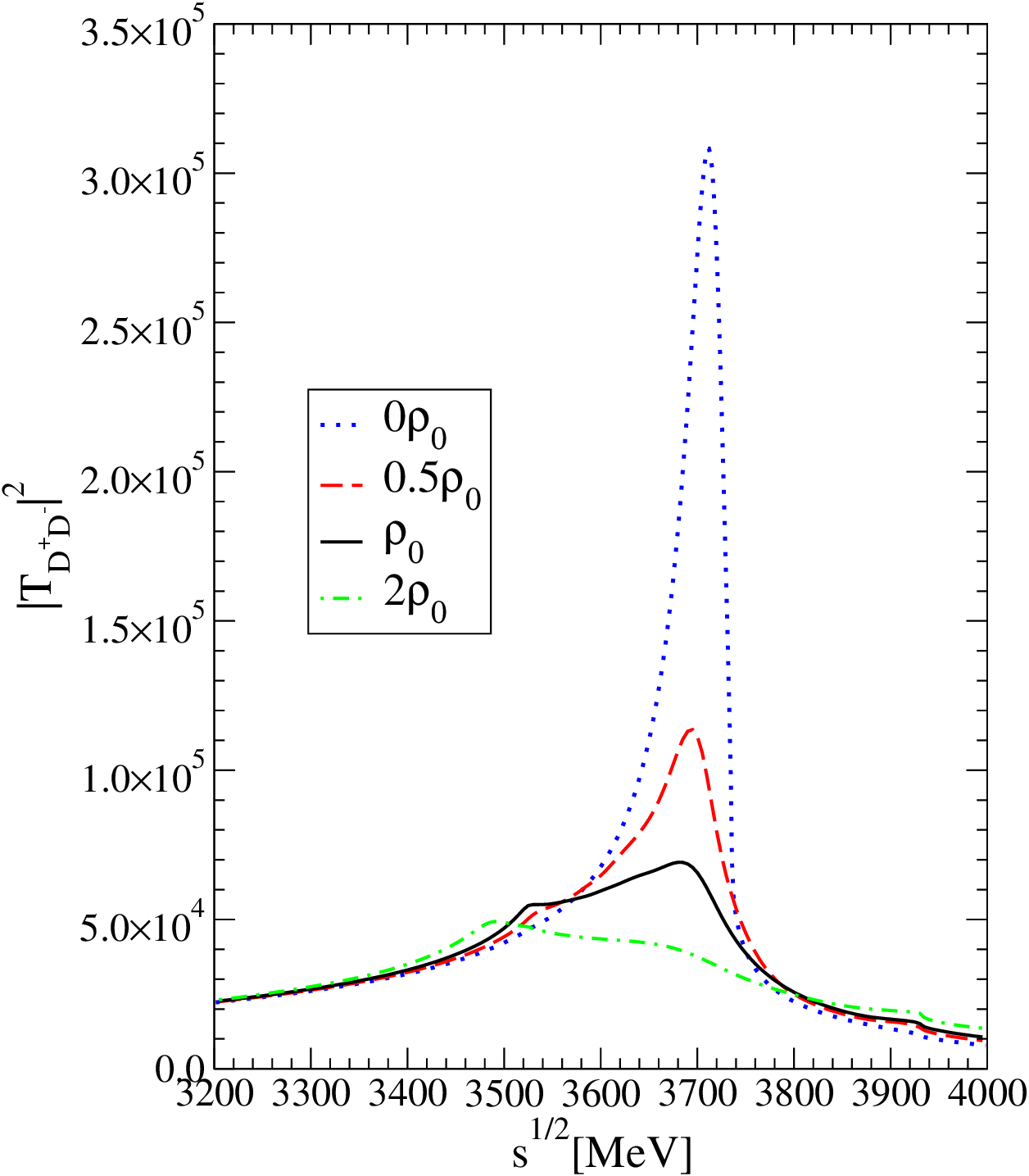}
\caption{$D_{s0}(2317)$ (left) and  $X(3700)$ (right) resonances in nuclear matter \label{fig4}}
\end{figure}


 In Fig.~\ref{fig4} the resonances $D_{s0}(2317)$ and $X(3700)$ are shown by displaying the squared transition amplitude for the corresponding dominant channel at different densities. The $D_{s0}(2317)$ and $X(3700)$ resonances, which have a zero and small width,
develop widths of the order of 100 and 200 MeV at normal nuclear matter density,  respectively. The origin can be traced back to the opening of new many-body decay channels as the $D$ meson gets absorbed in the nuclear medium via $DN$ and $DNN$ inelastic reactions. We do not extract any clear conclusion for the mass shift. We suggest to look at transparency ratios to investigate those in-medium widths. This magnitude, which gives the survival probability in production reactions in  nuclei, is very sensitive to the absorption rate of any resonance inside nuclei, i.e., to its in-medium width \cite{Hernandez:1992rv,Kaskulov:2006zc}.

\begin{table}
\begin{tabular}{c|cccc}
\hline
\hline 
  &   $^{12}$C    & $^{27}$Al &   $^{40}$Ca   &  $^{208}$Pb    \\
\hline 
1s&($-$8.51,$-$4.64)&($-$10.97,$-$4.38) &($-$11.12,$-$3.86)         &($-$11.54,$-$2.89) \\
1p&             &( $-$5.89,$-$5.98)&( $-$7.34,$-$4.97)  &($-$10.31,$-$3.20) \\
1d&             &                                            &                           &( $-$8.79,$-$3.63) \\
2s&             &              &                                                      &( $-$8.18,$-$3.80) \\
1f&             &              &                                                       &( $-$6.98,$-$4.25) \\
2p&             &              &                                                      &( $-$6.02,$-$4.53) \\
\hline
\hline
\end{tabular}
\caption{(B,$-\Gamma/2$) in MeV for $D^0-$nucleus bound states}
\label{table}
\end{table}

\section{D-mesic nuclei}
The possible formation of $D$-meson bound states in $^{208}$Pb was predicted \cite{tsushima99} relying upon a strong mass shift for $D$ mesons in the nuclear medium  based on a quark-meson coupling (QMC) model \cite{sibirtsev99}. The experimental observation of those bound states, though, might be problematic since, even if there are bound states, their widths could be very large compared to the
 separation of the levels. This is indeed the case for the potential derived from a SU(4) $t$-vector meson exchange model \cite{TOL07}.
However, the model that incorporates heavy-quark symmetry in the charm sector \cite{tolos09} generates widths of the $D$ meson in nuclear matter sufficiently small with respect to the mass shift to form bound states for $D$ mesons in nuclei.

In order to compute de $D$-nucleus bound states, we solve the Klein-Gordon equation (KGE). We concentrate on $D^0$-nucleus bound states since the coulomb interaction will prevent the formation of bound states for $D^+$ mesons. The potential that enters in the KGE is an energy-dependent one that results from the zero-momentum $D$-meson self-energy at the quasiparticle energy within the SU(8) model. In Table \ref{table} we show the binding energies ($B$) and widths ($-\Gamma/2$) of bound states of the $D^0$ meson in different nuclei. We observe that the $D^0$-nucleus states are weakly bound with significant widths \cite{carmen10} in contrast to previous results using the QMC model, in particular, for $^{208}$Pb \cite{tsushima99}.

\section{Conclusions and Outlook}

We have studied the properties of open-charm mesons in dense matter within a self-consistent coupled-channel approach taking, as bare interaction, different effective lagrangians. The in-medium solution  accounts for Pauli blocking effects and meson self-energies. We have analyzed the behaviour in this dense environment of dynamically-generated charmed baryonic resonances together with the evolution with density and temperature of the open-charm meson spectral functions. We have  discussed the implications of the properties of charmed mesons on the  $D_{s0}(2317)$ and the predicted $X(3700)$ in nuclear matter, and suggested to look at transparency ratios to investigate the changes in width of those resonances in nuclear matter. We have finally analyzed the possible formation of $D$-mesic nuclei. Only  
weakly bound $D^0$-nucleus states seem to be feasible within the SU(8) scheme that incorporates heavy-quark symmetry.

\begin{theacknowledgments}
L.T. acknowledges support from the RFF program of the University of Groningen. This work is partly supported by the EU contract No. MRTN-CT-2006-035482 (FLAVIAnet), by the contracts FIS2008-01661 and FIS2008-01143 from MICINN (Spain), by the Spanish Consolider-Ingenio 2010 Programme CPAN (CSD2007-00042), by the Generalitat de Catalunya contract 2009SGR-1289 and by Junta de Andaluc\'{\i}a under contract FQM225. We acknowledge the support of the European Community-Research Infrastructure Integrating Activity ``Study of Strongly Interacting Matter'' (HadronPhysics2, Grant Agreement n. 227431) under the 7th Framework Programme of EU.
\end{theacknowledgments}

\bibliographystyle{aipproc}

\begin{thebibliography}{9}

\bibitem{matsui86} T.~Matsui and H.~Satz, \emph{Phys. Lett. B} {\bf 178}, 416 (1986).

\bibitem{facility00} http://www.lepp.cornell.edu/Research/EPP/CLEO/; http://belle.kek.jp/; http://www.slac.stanford.edu/BF/

\bibitem{gsi00}http://www.gsi.de/~fair

\bibitem{LUT06} 
M.~F.~M.~Lutz, and C.~L.~Korpa, 
\emph{Phys.\ Lett. B} \textbf{633}, 43 (2006).

\bibitem{mizutani06} 
T.~Mizutani, and A.~Ramos, \emph{Phys.\ Rev. C}, {\bf 74}, 065201 (2006). 

\bibitem{TOL07}
L.~Tolos, A.~Ramos, and T.~Mizutani, 
\emph{Phys.\ Rev. C} {\bf 77}, 015207 (2008).


\bibitem{Tolos:2008di}
L.~Tolos, A.~Ramos, and A.~Polls, 
\emph{Phys. Rev. C} {\bf 65}, 054907 (2002);
L.~Tolos, A.~Ramos, and E.~Oset, 
\emph{Phys.\ Rev. C} {\bf 74}, 015203 (2006);
L.~Tolos, D.~Cabrera, and A.~Ramos, 
\emph{Phys.\ Rev. C}  {\bf 78}, 045205 (2008);

\bibitem{magas09}
C.~Garcia-Recio, V.~K.~Magas,T.~ Mizutani, J.~Nieves, A.~Ramos, L.~L.~Salcedo, and L.~Tolos,
\emph{Phys.\ Rev. D} {\bf 79}, 054004 (2009).

\bibitem{GarciaRecio:2005hy}
C.~Garcia-Recio, J.~Nieves, and L.~L.~Salcedo,
\emph{Phys.\ Rev. D} {\bf 74}, 034025 (2006). 

\bibitem{tolos09} 
L.~Tolos, C.~Garcia-Recio, and J.~Nieves,
\emph{Phys. Rev. C} {\bf 80}, 065202 (2009).

\bibitem{Amsler08}
C.~Amsler {\it et al.}  [Particle Data Group],
\emph{Phys.\ Lett. B} {\bf 667}, 1 (2008).

\bibitem{Kolomeitsev:2003ac} 
E.~E.~Kolomeitsev, and M.~F.~M.~Lutz,
\emph{Phys.\ Lett. B} {\bf 582}, 39 (2004).

\bibitem{guo06} 
F.~K.~Guo, P.~N.~Shen, H.~C.~Chiang, and R.~G.~Ping,
\emph{Phys.\ Lett. B} {\bf 641}, 278 (2006). 

\bibitem{Gamermann:2006nm} 
D.~Gamermann, E.~Oset, D.~Strottman, and M.~J.~Vicente Vacas,
\emph{Phys.\ Rev. D} {\bf 76}, 074016 (2007). 

\bibitem{Abe:2007sy} 
K.~Abe {\it et al.}  [Belle Collaboration],
\emph{Phys. Rev. Lett.}  {\bf 100}, 202001 (2008).

\bibitem{Gamermann:2007mu} 
D.~Gamermann, and E.~Oset, 
\emph{Eur. Phys. J. A} {\bf 36}, 189 (2008).  

\bibitem{Molina:2008nh} 
R.~Molina, D.~Gamermann, E.~Oset, and L.~Tolos, 
\emph{Eur. Phys. J. A} {\bf 42}, 31 (2009).

\bibitem{Hernandez:1992rv}
  E.~Hernandez, and E.~Oset,
  \emph{Z.\ Phys.\  A} {\bf 341}, 201 (1992).

\bibitem{Kaskulov:2006zc}
  M.~Kaskulov, E.~Hernandez, and E.~Oset,
  \emph{Eur.\ Phys.\ J.\  A} {\bf 31}, 245 (2007).

\bibitem{tsushima99}
K.~Tsushima, D.~H.~Lu, A.~W.~Thomas, K.~Saito, and R.~H.~Landau,
\emph{Phys. Rev. C} {\bf 59}, 2824 (1999).

\bibitem{sibirtsev99}
A.~Sibirtsev, K.~Tsushima  and A~.W.~Thomas, \emph{Eur. Phys. J. A} {\bf 6}, 351 (1999).

\bibitem{carmen10} C. Garcia-Recio, J. Nieves and L. Tolos, {\it in preparation}.

\end{thebibliography}

\end{document}